\documentclass[runningheads]{llncs}

\usepackage{caption}
\usepackage{subfigure}
\usepackage{graphicx}
\usepackage[cmex10]{amsmath}
\newtheorem{myDef}{Definition}

\begin{document}
\title{Differential Privacy in Privacy-Preserving Big Data and Learning: Challenge and Opportunity}
%
%\titlerunning{Abbreviated paper title}
% If the paper title is too long for the running head, you can set
% an abbreviated paper title here
%
\author{Honglu Jiang\inst{1}\orcidID{0000-0001-6014-0396} \and
Yifeng Gao\inst{1}\orcidID{0000-0002-0629-050X}
\and S M Sarwar \inst{1}\orcidID{0000-0003-0772-8064} \and Luis GarzaPerez\orcidID{ 0000-0001-9482-709X} \inst{1} \and  Mahmudul Robin \orcidID{ 0000-0002-3157-2189} \inst{1}}
\authorrunning{H. Jiang et al.}
% First names are abbreviated in the running head.
% If there are more than two authors, 'et al.' is used.
%
\institute{The University of Texas Rio Grande Valley, Edinburg, TX, 78504, USA
\email{honglu.jiang, yifeng.gao, sm.sarwar01,luis.garzaperez,mahmudul.robin01@utrgv.edu}\\
}
\maketitle              % typeset the header of the contribution
\begin{abstract}
Differential privacy (DP) has become the de facto standard of privacy preservation due to its strong protection and sound mathematical foundation, which is widely adopted in different applications such as big data analysis, graph data process, machine learning, deep learning, and federated learning. Although DP has become an active and influential area, it is not the best remedy for all privacy problems in different scenarios. Moreover, there are also some misunderstanding, misuse, and great challenges of DP in specific applications. In this paper, we point out a series of limits and open challenges of corresponding research areas. Besides, we offer potentially new insights and avenues on combining differential privacy with other effective dimension reduction techniques and secure multiparty computing to clearly define various privacy models.

\keywords{Differential privacy \and Deep learning \and Big data.}
\end{abstract}
\section{Introduction}

Organizations, companies and governments collect data from a variety of sources, including social networking, transactions, smart Internet of Things devices, industrial equipment, electronics commercial activities, and more, which can be used to dig out valuable information hidden behind the massive data for modern life. 
The extensive collection and further processing of personal information in the context of big data analytics and machine learning-based artificial intelligence results in serious privacy concerns. For example, in March 2018, Facebook-Cambridge Analytica was reported to use the personal data of millions of people's Facebook profiles harvested without their consents for political advertising purposes in the 2016 US presidential election, which was a great political scandal and caused an uproar in the world. Despite the benefits of analytics, it cannot be accepted that big data comes at a cost for privacy. Therefore, the present study shifts the discussion from ``big data versus privacy'' to ``big data with privacy'', adopting the privacy and data protection principles as an essential value \cite{DGD}. Privacy-preserving data publishing (PPDP) and various artificial intelligence-empowered learning/computing have gained significant attentions in both academia and industry. It is, thus, of utmost importance to craft the right balance between making use of big data technologies and protecting individuals’ privacy and personal data \cite{DGD}.

Intuitively, one can make use of the simple na{i}ve identity removal to protect data privacy, but in practice, it does not always work. For instances, AOL released an anonymized partial three-month search history to the public in 2006. Although personally identifiable information was carefully processed, some identities were accurately reidentified. For example, \emph{The New York Times} immediately located the following individual: the person with number $4417749$ was a $62$-year-old widowed woman who suffered from some diseases and has three dogs. Such real-world privacy leakage problems and attack instances clearly demonstrate the importance of data privacy preservation.

The problem of data privacy protection was first put forward by Dalenius in the late 1970s \cite{DT} --- Dalenius pointed out that the purpose of protecting private information in a database is to prevent any user (including legitimate users and potential attackers) from obtaining accurate information about arbitrary individuals. Following that, many privacy preservation models with strong operability including $k$-anonymity, $l$-diversity \cite{MKV}, $t$-closeness \cite{LLV} were proposed. However, each model generally provides protection against only a specific type of attacks and cannot defend against newly developed ones. A fundamental cause of this deficiency lies in that the security of a privacy preservation model is highly related to the background knowledge of an attacker. Nevertheless, it is almost impossible to define the complete set of possible background knowledge an attacker may have.

Dwork originally proposed the concept of \emph{differential privacy} (DP) to protect against the privacy disclosure of statistical databases in 2006 \cite{DCD}. Under differential privacy, query results of a dataset are insensitive to the change of a single record. That is, whether a single record exists in the dataset has little effect on the output distribution of the analytical results. As a result, an attacker cannot obtain accurate individual information by observing the results since the risk of privacy disclosure generated by adding or deleting a single record is kept within an acceptable range. Unlike anonymization model, DP makes the assumption that an attacker has the maximum background knowledge, which rests on a sound mathematical foundation with a formal definition and rigorous proof.

It is worth noting that differential privacy is a definition or standard for quantifying privacy risks rather than a single tool, which is widely used in statistical estimations, data publishing, data mining, and machine learning. It is a new and promising privacy framework and has become a popular research topic in both academia and industry, which can be potentially implemented in various application scenarios. However, DP is a strict privacy standard, the data utility is likely to be poor while providing a meaningful privacy guarantee. The goal of this paper is to summarize and analyze the state-of-the-art research and investigations in the field of differential privacy and its applications in privacy-preserving data publishing, machine learning, deep learning, and federated learning, to point out a series of limits and open challenges of corresponding research areas, so as to provide some approachable strategies for researchers and engineers to implement DP in real world applications. In our paper, we place more focus on practical applications of differential privacy rather than detailed theoretical analysis of differentially private algorithms.

The rest of this paper is organized as follows. We present the background knowledge of differential privacy in Section \ref{pre}. Section \ref{data} introduces differentially private data publishing problem and presents some challenges on this problem. In Section \ref{mac}, we summarize existing research on the application of differential privacy to deep learning and federated learning. Section \ref{con} concludes the paper with some future research discussion and open problems on differential privacy applications.

\section{Preliminary of Differential Privacy}\label{pre}

Differential privacy can be achieved by injecting a controlled level of statistical noise into a query result to hide the consequence of adding or removing an arbitrary individual from a dataset. That is, when querying two almost identical datasets (differing by only one record), the results are differentially privatized in that an attacker cannot glean any new knowledge about an individual with a high degree of probability, i.e., whether or not a given individual is present in the dataset cannot be guessed. 

\subsection{Definition of Differential Privacy}
\begin{figure}[t]
 \centering
\includegraphics[width=0.7\textwidth]{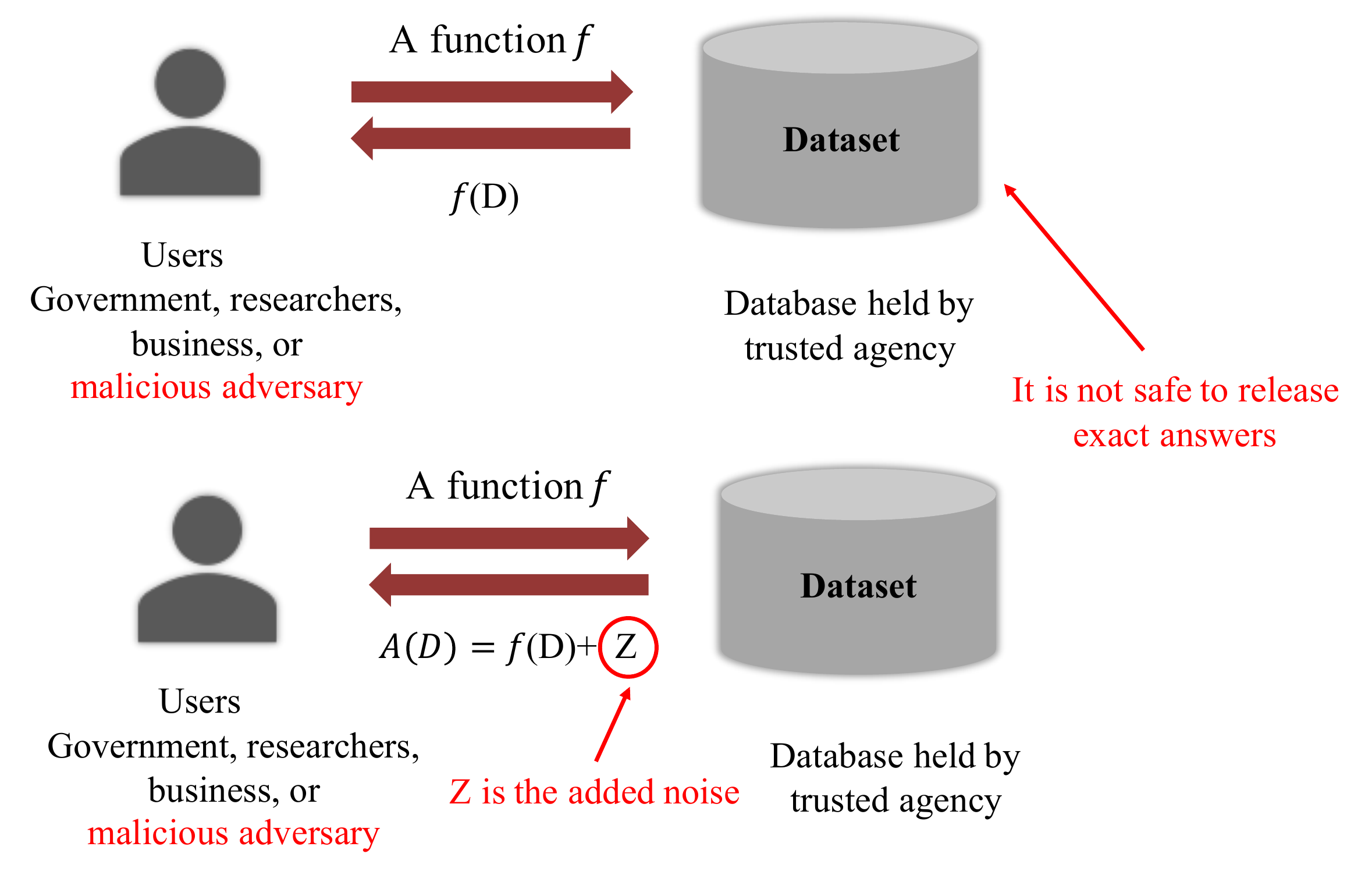}
 \caption{The framework of differential privacy}
 \label{fig:frame}
\end{figure}

Let $f$ be a query function to be evaluated on a dataset $D$. Algorithm $A$ runs on the dataset $D$ and sends back $A(D)$. $A(D)$ could be $f(D)$ with a controlled amount of random noise added. The goal of differential privacy is to make $A(D)$ as much close to $f(D)$ as possible, thus ensuring data utility (enabling the user to learn the target value as accurately as possible), while preserving the privacy of the individuals with the added random noise. The main procedure can be seen in Figure \ref{fig:frame}.

\begin{myDef}
(Neighboring Datasets) Two datasets $D$ and $D'$ are considered to be neighboring ones if $d(D,D')=1$, where $d(D,D')$ is the number of records $D$ and $D'$ differ. %, i.e. $d(D,D')=|\{i: D_i\neq D'_i\}|$.
\end{myDef}

\begin{myDef} \label{def:DP}
(Differential Privacy \cite{DMN})  A randomized algorithm $A$ is $(\epsilon, \delta)$-differentially private if for any two datasets $D$ and $D'$ with $d(D,D')=1$, and for all sets $S$ of possible outputs, we have
\begin{equation}
Pr[A(D)\in S]\leq e^{\epsilon}Pr[A(D')\in S]+\delta,\nonumber
\end{equation}
\end{myDef}
where $\epsilon$ and $\delta$ are non-negative real numbers.

When $\delta=0$, the algorithm becomes $\epsilon$-differentially private. We say a mechanism gives $\delta$-approximate differential privacy when $\delta\neq0$. The $\epsilon$ is often a small positive real number called \emph{privacy budget}, which is used to control the probability of the algorithm $A$ getting almost the same outputs from two neighboring datasets. It reflects the level of privacy preservation that algorithm $A$ can provide. For example, if we set $\epsilon=\ln2$, the result $S$ is at most twice as likely to be generated by dataset $D$ as by any of $D$'s neighbor $D'$.

The smaller the $\epsilon$, the higher the level of privacy preservation. A smaller $\epsilon$ provides greater privacy preservation at the cost of lower data accuracy with more additional noise. When $\epsilon=0$, the level of privacy preservation reaches the maximum, i.e., ``perfect'' protection. In this case, the algorithm outputs two results with indistinguishable distributions but the corresponding results do not reflect any useful information about the dataset. Therefore, the setting of $\epsilon$ should consider the trade-off between privacy requirements and data utility. In practical applications, $\epsilon$ usually takes very small values such as $0.01$, $0.1$, or $\ln2$, $\ln3$. 

\subsection{Noise Mechanism of Differential Privacy}

Sensitivity is the key parameter to determine the magnitude of the added noise, that is, the largest change to the query result caused by adding/deleting any record in the dataset. Accordingly, global sensitivity, local sensitivity, smoothing upper bound, and smoothing sensitivity are defined under the differential privacy model. Because of the limitation of space, we will specifically introduce them here.

\textbf{(1) Laplace Mechanism}

The Laplace distribution (centered at $\mu$) with scale $b$ is the distribution with probability density function
$$h(z)=\frac{1}{2b}\exp(-\frac{|z-\mu|}{b}).$$
Let $Lap(b)$ denote the Laplace distribution (centered at $0$) with scale $b$.

\begin{myDef}
(Laplace Mechanism \cite{DMN}) For dataset $D$ and function $f: D\rightarrow R^{d}$ with global sensitivity $GS_{f}$, the Laplace mechanism
 $A(D)=f(D)+Z$ is $\epsilon$-differentially private, 
 where $Z\sim Lap(GS_{f}/\epsilon)$.
\end{myDef}

The Laplace mechanism is suitable for the protection of numerical results. Taking an example Laplace mechanism for the counting function, since the global sensitivity of counting is $1$, that is $GS_{f}=1$, if we choose $\epsilon=0.1$, the Laplace mechanism outputs $3+Lap(10)$.

\textbf{(2)Exponential Mechanism}

The Laplace mechanism is appropriate only for preserving the privacy of numerical results. Nevertheless, in many practical implementations, query results are entity objects. McSherry \emph{et al.} put forward the exponential mechanism \cite{MTK} for the situations where the ``best'' needs to be selected. Let the output domain of a query function be $Range$, and each value $r\in Range$ be an entity object. In the exponential mechanism, the function $q(D,r)$, which is called the \emph{ utility function} of the output value $r$, is employed to evaluate the quality of $r$.

\begin{myDef}
(Exponential Mechanism \cite{MTK}) Given a random algorithm $A$ with the input dataset $D$ and the output entity object $r\in Range$, let $q(D,r)$ be the utility function and $\Delta q$ be the global sensitivity of function $q(D,r)$. If algorithm $A$ selects and outputs $r$ from $Range$ at a probability proportional to $\exp(\frac{\epsilon q(D,r)}{2\Delta q})$, then $A$ is $\epsilon$-differentially private.
\end{myDef}

\subsection{Local Differential Privacy}\label{sec:ldp}
Traditional centralized differential privacy provides privacy protection based on a premise that there is a trusted third-party data collector who does not steal or disclose user's sensitive information, while local differential privacy \cite{DJW} does not assume the existence of any trusted third-party data collector. Instead, it transfers the process of data privacy protection to each user, making each user independently deal with and protect personal sensitive information.

\begin{myDef}
\label{ldp}
(Local Differential Privacy \cite{DJW}) Given $n$ users, with each corresponding to a record. A privacy algorithm $M$ with definition domains $Dom (M)$ and $Ran (M)$ satisfies the $\epsilon$-local differential privacy if $M$ obtains the same output result $t^{*}$ $(t^{*}\subseteq Ran(M))$ on any two records $t$ and $t'$ $(t, t'\in Dom(M))$:
$$Pr[M(t)=t^{*}]\leq e^{\epsilon}\times Pr[M(t')=t^{*}]$$
\end{myDef}

 \begin{figure}[t]
 \centering
\includegraphics[width=0.7\textwidth]{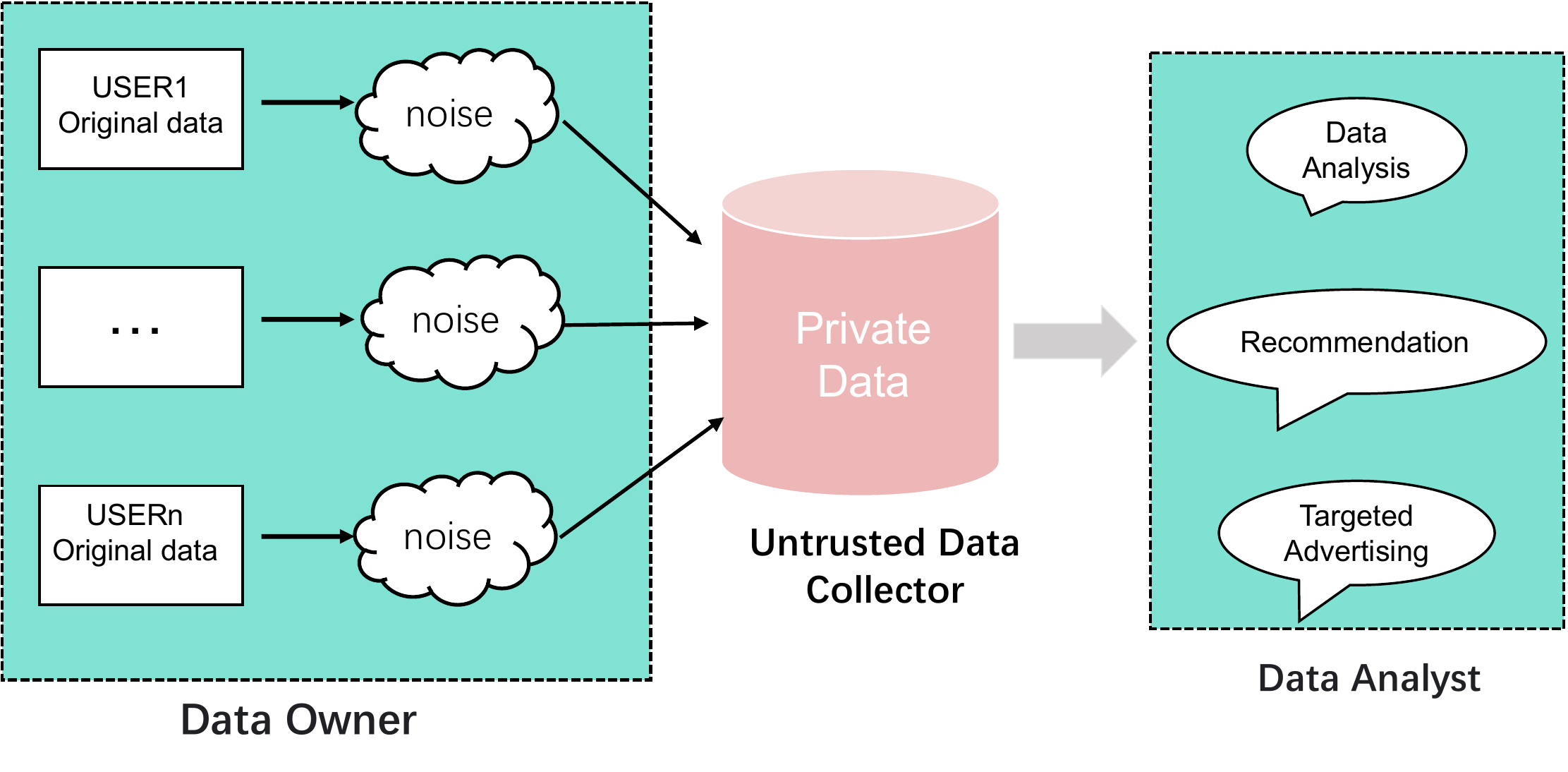}
 \caption{The framework of local differential privacy}
 \label{fig2}
       \vspace{-5mm}
\end{figure}

One can see from this definition that local differential privacy provides privacy by controlling the similarity between the output results of any two records, while each user processes its individual data independently, that is, the privacy preserving process is transferred to a single user from the data collector, such that a trusted third party is no longer needed and privacy attacks brought from the data collection of untrusted third-party is thus avoided. The framework of local differential privacy can be seen in Figure \ref{fig2}.

\section{Differentially Private Data Publishing}\label{data}

\subsection{Differential Privacy in Tabular Data Publishing}

The goal of differentially private data publishing is to output aggregate/synthetic information to public without disclosing any individual's information. Generally, there are two settings in the data publishing scenario, interactive and non-interactive. In the first setting, users make queries request to the data curator, who answers the query with a noisy result. The fixed privacy budget will be exhausted as the number of queries increases. In the non-interactive setting, the data curator publishes statistical information related to the dataset that satisfies differential privacy. When the queries are submitted, the corresponding query result is directly returned from the published synthetic dataset.

The challenge of interactive setting is that the number of queries is limited while the privacy budget $\epsilon$ is easily exhausted. That is, a higher accuracy result for one query with less noise results and a larger $\epsilon$ usually results in a smaller number of queries. 

High sensitivity presents a big challenge on the data publishing in the non-interactive setting, while high sensitivity means large magnitude of noise and low data utility especially for big data and complex data, which we will detailed introduce in Section \ref{sec:cha}. Another problem is that the published synthetic dataset can only be used for particular purposes or targeted a fixed query function.

  \subsection{Differential Privacy in Graph Data Publishing}
    
With the widespread application of social networks, the increasing volumes of user-generated data have become a rich source which can be published to third parties for data analysis and recommendation system. Generally, social networking data can be modeled as graph $G(V,E)$, where $V$ is a set of nodes and $E$ is a set of relational activities between nodes. Analyzing graph data such as analysis of social network data has great potential social benefits and help generate insights into the laws of data change and trend characteristics. 
Most popular tasks of social network analysis include degree distribution, subgraph counting (triangle
counting, k-star counting, k-triangle counting,etc.) and edge weight analysis. In reality, various types of privacy attacks such as de-anonymization attacks \cite{SWN,JTJ,JYH,SGE,JYC}, inference attacks \cite{GNT2014,LSW2013} on social networks have raised the stakes for privacy protection while a large amount of personal user data have been exposed.

\begin{figure}[t]
 \centering
\begin{minipage}{0.4\linewidth}
  \centerline{\includegraphics[width=4.0cm]{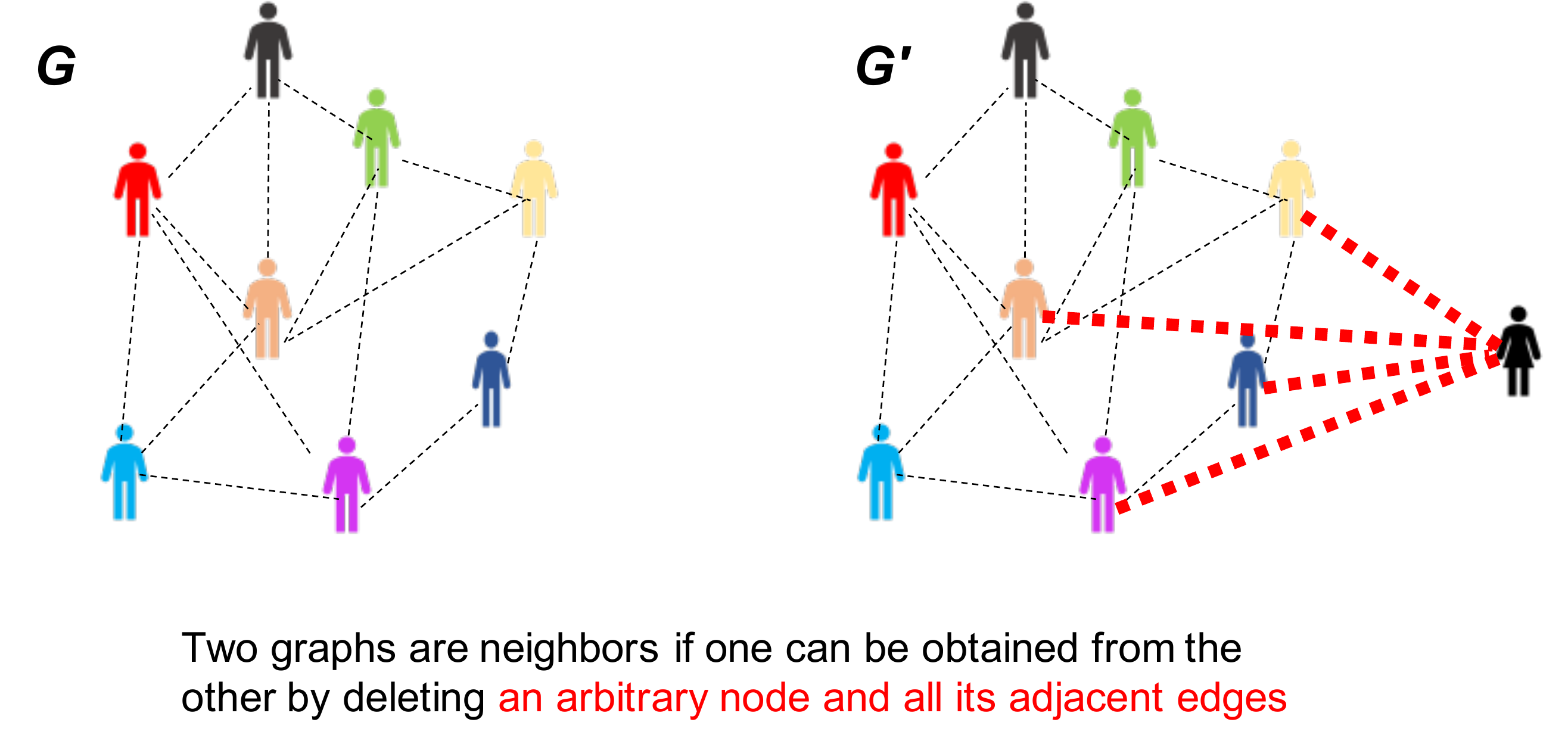}}
  \centerline{(a) Node differential privacy}
\end{minipage}
\hfill
\begin{minipage}{0.4\linewidth}
  \centerline{\includegraphics[width=4.0cm]{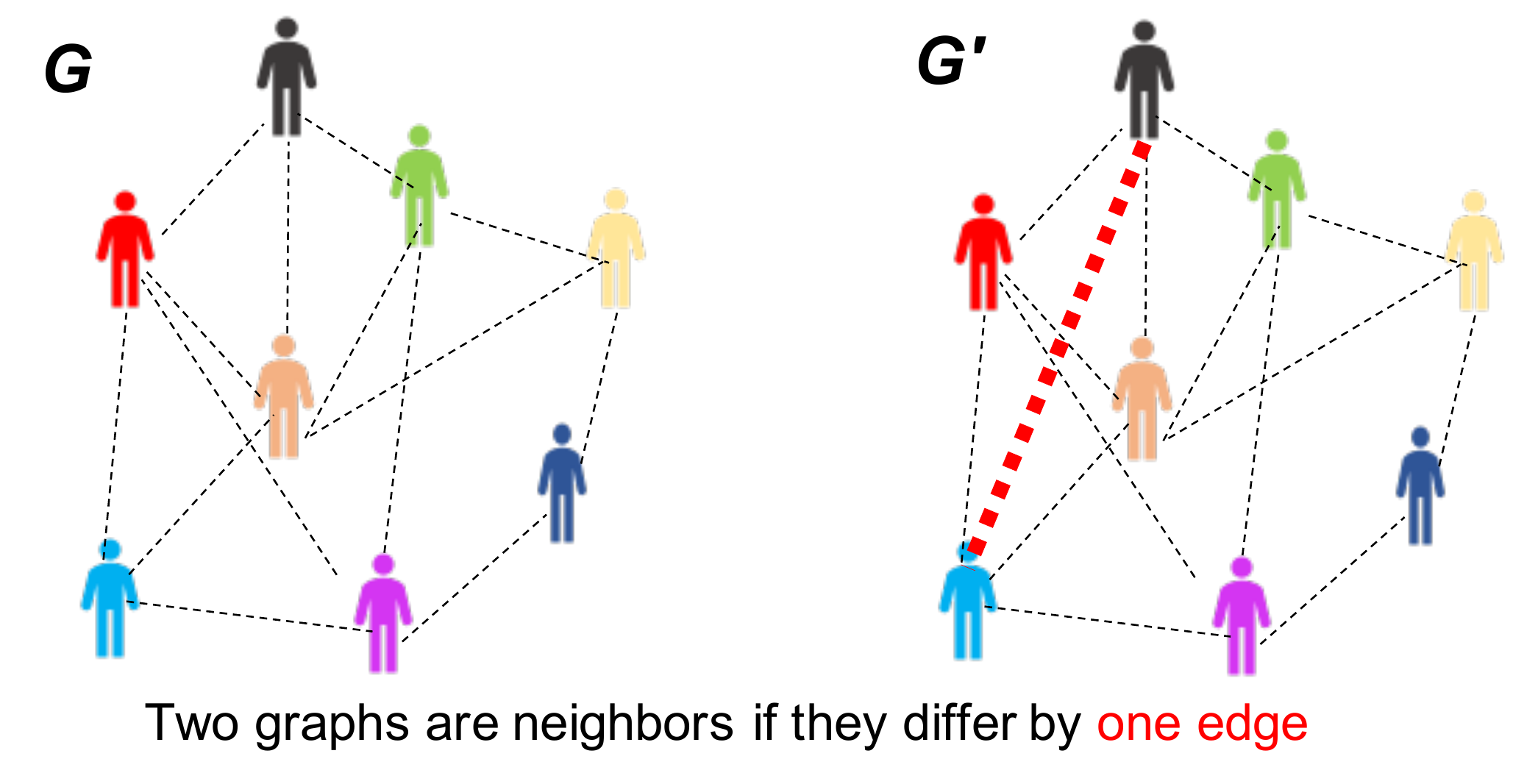}}
  \centerline{(b) Edge differential privacy}
\end{minipage}

\vfill

\begin{minipage}{0.4\linewidth}
  \centerline{\includegraphics[width=4.0cm]{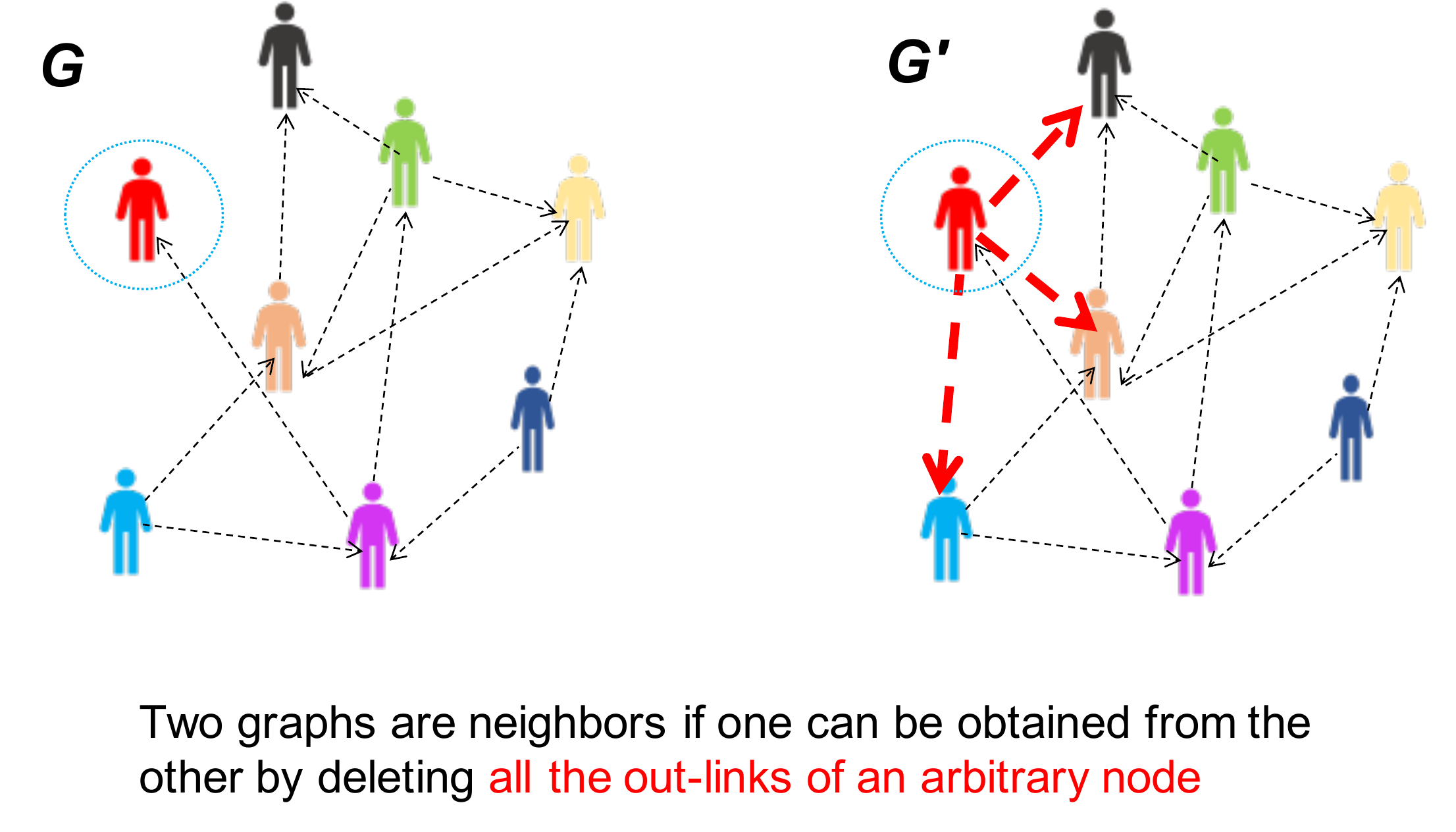}}
  \centerline{(c) Outlink differential privacy}
\end{minipage}
\hfill
\begin{minipage}{0.4\linewidth}
  \centerline{\includegraphics[width=4.0cm]{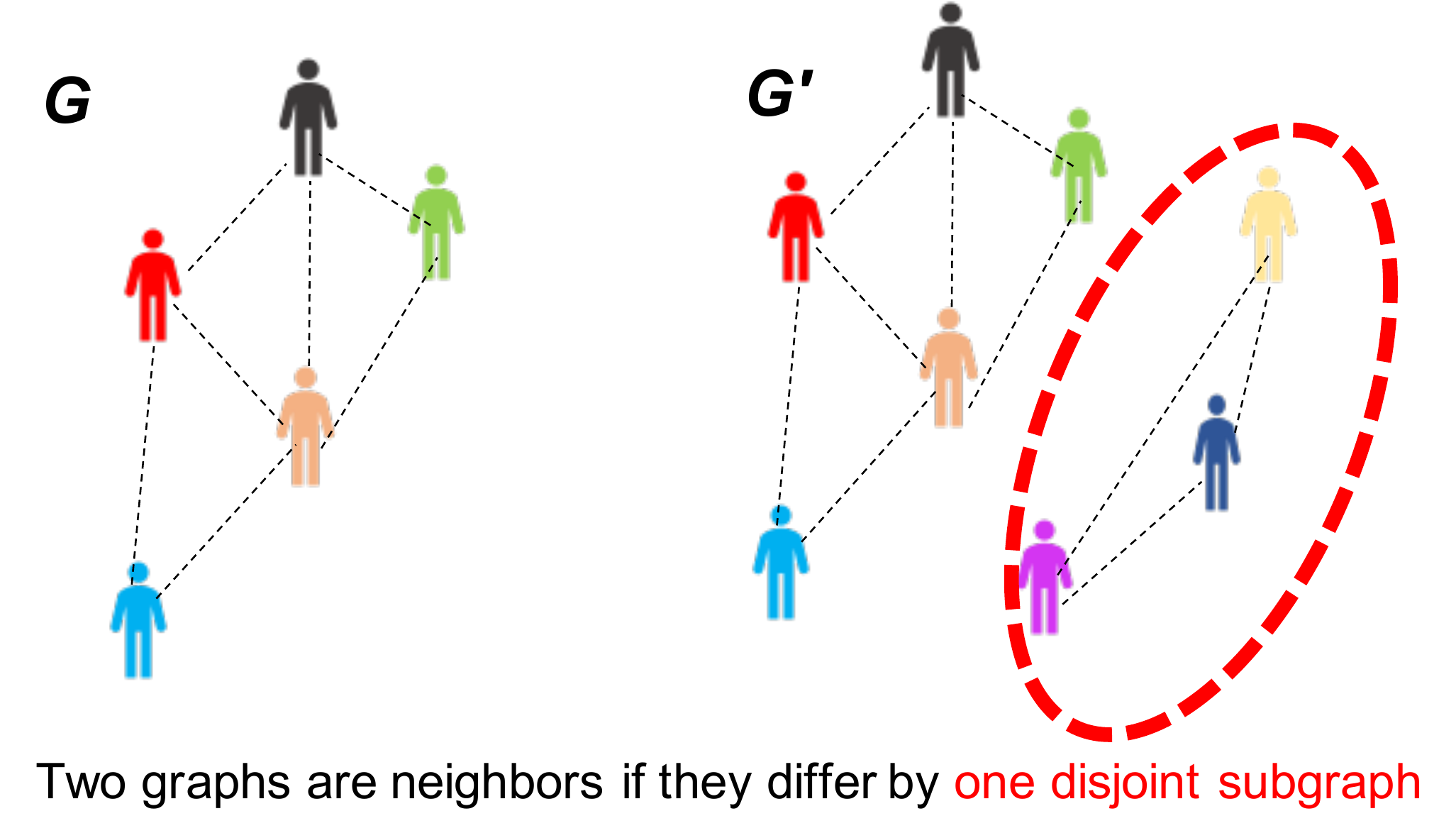}}
  \centerline{(d) Partition differential privacy}
\end{minipage}
  \caption{Differential privacy definitions in graph data}  
   \label{fig1}
    \vspace{-5mm}

% \begin{subfigure}
%\begin{subfigure}[b]{0.4\textwidth}
 % \includegraphics[width=\textwidth]{./Figures/1}
%\caption{Node differential privacy}
% \end{subfigure}
%\begin{subfigure}[b]{0.4\textwidth}
%  \includegraphics[width=\textwidth]{./Figures/2}
% \caption{Edge differential privacy}
%\end{subfigure}
    
%\begin{subfigure}[b]{0.4\textwidth}
 %\includegraphics[width=\textwidth]{./Figures/3}
 %\caption{Outlink differential privacy}
%\end{subfigure}
%     \begin{subfigure}[b]{0.4\textwidth}
 %\includegraphics[width=\textwidth]{./Figures/4}
 %  \caption{Partition differential privacy}
 %    \end{subfigure}     
 %   \caption{Differential privacy definitions in graph data}  
  %    \label{fig1}
  %    \vspace{-5mm}
\end{figure}

However, the privacy issue of graph is more complicated starting from how to model and formalize the notion of ``privacy" in graph network. Differential privacy originates from tabular data, while the key to extending differential privacy to social networks is to determine the neighboring input entries, that is, how to define ``adjacent graphs”. Figure \ref{fig1} shows existing definitions of DP in graph data, namely, node differential privacy, edge differential privacy, outlink differential privacy, partition differential privacy; detailed information can be referred to \cite{JPY}.
 
 \subsection{Challenges on Differentially Private Data Publishing.} \label{sec:cha}
 
 In this subsection, we present a few challenges and open problems on differentially private data publishing especially for big data, complex network, dynamic and continuous data publishing.
 
 As what it reads, big data deal with massive amounts of data at a great speed passing, which exhibit various characteristics that cover challenges like gathering, analysis, storage and privacy preservation. Of the many characteristics of big data, \emph{5V} characterizes big data’s nature the best, namely Volume, Velocity, Variety, Veracity and Value. 
 
\noindent\textbf{Differential privacy on complex and high volume network structure.}
Network structures such as social networks and traffic networks are often complex. Since query sensitivities are usually high, much noise has to be added to query results to achieve differential privacy. Nevertheless, the noise may significantly affect the output data utility, resulting in useless data. Moreover, it may
be hard to effectively compute sensitivities, either global or smooth, precise or approximate, as the computational complexity may be too high (or even NP-hard) to be practical for many complex graph network analysis queries. 
 
\noindent\textbf{Differential privacy on high dimensional Data.}
Most differentially private data publishing techniques cannot work effectively for high dimensional data. On one hand, since the sensitivities and entropy of different dimensions vary, evenly distributing the total privacy budget to each dimension degrades the performance. Moreover, ``Curse of Dimensionality" is the common challenge in big data perturbation which means a dataset contains high dimensions and large domains resulting in a pretty low ``Signal-to-Noise" and extremely low data utility even useless.

\noindent\textbf{Differential privacy on correlated data.}
Differential privacy offers a neat privacy guarantee while it is a strict privacy standard, while assumes all the data are independent, while the correlation or dependence may undermine the privacy guarantees of differential privacy mechanisms. Unfortunately, the real-world gathered data can not be strictly independent, which is not only tuple (record) correlated but also attribute information correlated. For example, the salary information in strongly correlated with education level and occupation in a dataset.

\noindent\textbf{Differential privacy on high-velocity data.}
Velocity in big data refers to the crucial characteristic of capturing data dynamically. In practical applications, the data are dynamically updated such as recommendation system, trajectory data to capture the evolutionary behaviors of various users. Differential privacy on continuous flow of data faces critical challenges of great noise accumulation and privacy budget allocation for each time sequence.

\section{Differentially Private Machine Learning}\label{mac}

\subsection{Differential Privacy in Deep Learning} 
 
 The privacy protection provided by DP also could benefit the existing deep learning model. The Differential Privacy framework for deep learning is illustrated in Fig. 4. Generally, the noise can be added into the gradient, input, and embedding. Adadi et al. \cite{abadi2016deep} introduce the first DP preserved optimization algorithm named DPSGD. The DP is achieved by adding Gaussian noise in every SGD optimization step. Arachchige et al. \cite{arachchige2019local} introduce a model named LATENT. The framework achieves the protection by transferring the real-vale low dimensional representation into a discrete vector. Lecuyer et al. \cite{lecuyer2019certified} proposed a model named PixelDP. The framework achieves the goal by adding Gaussian noise in the hidden layers of a CNN model. Different from these works, Phan et al.\cite{phan2017adaptive} proposed a method that directly manipulates the inputs. The model induces different levels of noise for each pixel of an image based on a relevant score\cite{bach2015pixel}.

 \subsection{Differential Privacy in Federated Learning}

The research field of Federated Learning focuses on learning a model where data is stored in a distributed system. As pointed out by Wei et al. \cite{wei2020framework}, attackers can retrieval the data information through the gradient, a DP preserved learning model could protect such information leakage in the Federated Learning setting. Wei et al. \cite{wei2020federated} integrated DP algorithm into the Secure Multiparty Computation(SMC) framework. DP is used to encrypt the response for each query in the SMC. Geyer et al. \cite{geyer2017differentially} introduce a DP algorithm focusing on removing the data source info. In addition to using the same SGD algorithm framework as DPSGD, the algorithm also will randomly ignore a portion of the data to protect data privacy.

\subsection{Challenges on Differentially Private Machine learning}

\noindent\textbf{Model Dependency.} Other than the gradient-based approach, most deep-learning based DP algorithms introduced in this paper are highly related to the deep learning model. For example, LATENT and PixelDP are designed only for CNN. A DP approach that does not rely on the data and model could be promising research in the DP research field.

\noindent\textbf{Accuracy loss of federated learning due to added noise.} In federated learning model, differential privacy-based approaches add noise to the uploaded parameters which will degrade the model accuracy inevitably and further affect the convergence of the global aggregation. Moreover, there are few results about practical frameworks integrating differential privacy and other cryptography-based methods, which hinders the industrial development of federated learning.

\section{Future directions and Conclusions}\label{con} 

Differential privacy is a strong standard of privacy protection with a solid mathematical definition which can be applied in various application scenarios, however differential privacy is not a panacea for all privacy problems and the research on differential privacy is still in its infancy stage. There are still some misunderstandings, inappropriate applications and flawed implementations in differential privacy. In this section, we propose a few future research problems and open problems that worthy of more attention.

 \subsection{Combination of Differential Privacy and Other Technologies}
 
 As we mentioned about the privacy preservation of high dimensional data, it is feasible and promising to combine effective dimensionality reduction techniques with differential privacy to address this issue. Specifically, it is possible to try both linear and non-linear transformation such as compressive sensing and manifold learning which maps a high-dimensional space to a low-dimensional representation.

With the great high privacy concern on Federated learning, IoT network and other distributed environment, the combination of local differential privacy, multiparty computations and sampling and anonymization will be a future topic which needs open-ended exploration. Secure multiparty computation is a type of cryptography-based which could be concerning and infeasible on computationally constrained devices, while anonymization model has its own shortcomings about the assumption on background knowledge. However, the combination of these techniques can boost the performance of differential privacy. Specifically, differential privacy with a sampling processing can greatly amplify the privacy preservation level \cite{NWQ}, based on which we can adapt the idea of anonymization to participants of DP processing. For example, in the scenario of federated learning, we can randomly pick up the clients and parts of differentially private local updates to form a shuffle model. Moreover, inter-discipline techniques between local differential privacy and secure multiparty computation involve the secure computation, privacy preservation and dataset partition, which need to tackle with the high communication cost and low data utility.

 \subsection{Variation of Differential Privacy and Personalized Privacy}
 
 Differential privacy provides strong and strict privacy guarantee at the cost of low data utility while it may be too strong and not necessary in some practical applications. To achieve a better tradeoff between privacy and preservation, various relax and extensions of differential privacy need to be proposed and in fact many of these definition have been proposed such as \emph{crowd-blending privacy},\emph{individual differential privacy}, and \emph{probabilistic indistinguishability}. However, most of these are still in the stage of theoretical definition or be specific scenarios. The great challenge is that how to widely apply to these extensions to practical applications.
 
 On the other hand, conventional private data privacy preservation mechanisms aim to retain as much data utility as possible while ensuring sufficient privacy protection on sensitive data while such schemes implicitly assume that all data users have the same data access privilege levels. Actually, data users often have different levels of access to the same data, personalized requirements of privacy preservation level or data utility. It is a big challenge to achieve personalized privacy and multi-level data utility while the uniform framework itself is a hard problem.
       \vspace{-3mm}
\subsection{Misunderstandings of Differential Privacy vs More Than Privacy}

As we mentioned in differentially private data publishing, the data utility of outputs are likely to be very poor or with large privacy budget, that is lower privacy preservation level, which we cannot sure how much privacy it can provides. Moreover, when differential privacy is applied to federated learning, it is used on local updates of parameters while traditional differential privacy is designed for record data contributed by different individuals on the basis of assumption that the data are independent. However, in federated/distributed learning, all local data are from the same client which have little possibility to be independent.

In contrast, differential privacy can do more while there exists misconceptions and misuse of differential privacy. Besides providing privacy preservation through hiding individual information in the aggregate
information, from the opposite perspective of its definition, differential privacy can ensure that the
probability of outcomes unchanged when modifying any individual record in the training data, and the application of this property needs to be explored. Secondly, differential privacy can also protect against the malicious attacks in learning techniques such as poisonous attacks in federated learning which can help improve the accuracy of training model. Thirdly, specific differentially private methods can be combined with reward mechanisms in distributed learning to provide privacy preservation and incentivize more clients to participate in the learning process at the same time.
      \vspace{-4mm}
%
% ---- Bibliography ----
%
% BibTeX users should specify bibliography style 'splncs04'.
% References will then be sorted and formatted in the correct style.
%
\bibliographystyle{splncs04}
\bibliography{ref}
\end{document}